\RequirePackage{fix-cm}
\documentclass[smallextended]{svjour3}       
\smartqed  
\usepackage{graphicx}
\usepackage{latexsym}
\usepackage{epsfig,latexsym,amssymb,amsmath}
\begin{document}

\title{Black Hole Interior Mass Formula       
}


\author{Parthapratim Pradhan}
\institute{\at Department of Physics\\
           Vivekananda Satabarshiki Mahavidyalaya \\
           Manikpara, West Midnapur\\
            West Bengal-721513, India. \\
            \email{pppradhan77@gmail.com}}
\date{Received: date / Revised version: date}

\maketitle

\begin{abstract}
We argue by explicit computations that, although the area product, horizon radii product, entropy product 
and \emph {irreducible mass product} of the event horizon and Cauchy horizon  are universal, the 
\emph{surface gravity product}, \emph{surface temperature product} and \emph{Komar energy product} of 
the said horizons do not seem to be universal for Kerr-Newman (KN) black hole space-time. We show the 
black hole mass formula on the  \emph{Cauchy horizon} following the seminal work by Smarr\cite{smarr} for
the outer horizon. We also prescribed the \emph{four} laws of black hole mechanics for the 
\emph{inner horizon}. New definition of the extremal limit of a black hole is discussed.
\end{abstract}

\section{Introduction}
The intriguing features of  a stationary axially symmetric  black hole is that  the product of the horizon areas are often 
be independent of the mass of the black hole space-time. Rather such products  depend on the charge and angular
momentum of the black hole space-time. They may also be  formulated in terms of the proper radii of the 
Cauchy horizon (Cauchy horizon) and event horizon (event horizon).

It is also known that every regular axially symmetric  and stationary space-time of
Einstein-Maxwell system with surrounding matter has a regular Cauchy horizon (${\cal H}^-$) inside the event
horizon (${\cal H}^+$) if and only if the  angular momentum $J$
and charge $Q$ do not both vanish. In contrast, the Cauchy horizon becomes singular and
approaches a curvature singularity in the limit $J\rightarrow 0$, $Q \rightarrow 0$ \cite{ansorg,ansorg1,ansorg2}.

The presence of the Cauchy horizon implies that in Boyer-Lindquist coordinates, the stationary and axi-symmetric 
Einstein-Maxwell electro-vacuum  equations are hyperbolic in nature with in the interior vicinity of ${\cal H}^+$.
The two horizons ${\cal H}^+$ and ${\cal H}^-$  are described by the future and past boundary of this hyperbolic 
region. Remarkably, if the inner  Cauchy horizon exists (i.e. if $J$ and $Q$ do not vanish simultaneously), then
the product of the  area $A_{\pm}$ of the horizons ${\cal H}^\pm$ for KN family are expressed  by the
relation \cite{ansorg,ansorg1,ansorg2}
\begin{eqnarray}
{\cal A}_{+} {\cal A}_{-} &=& (8\pi)^2\left(J^2+\frac{Q^4}{4}\right) ~.\label{proarKN}
\end{eqnarray}
which is remarkably independent of the mass $({\cal M})$. $J$ and $Q$ are the angular momentum
and charge of the black hole, respectively.

From the various string theoretic models and holographic principle followed by an observation 
suggests that the products of the certain Killing horizon areas are in fact independent of the 
black hole mass. From the idea of statistical mechanics based on microscopic models counting BPS 
states determined that this product of areas are sometimes quantized. Thus in the super-symmetric
extremal limit, one obtains \cite{larsen,cvetic,cvetic1,chen,castro,visser1,val}
\begin{eqnarray}
{\cal A}_{+} {\cal A}_{-} &=& \left(8\pi{\ell _{pl}}^2\right)^2 N , \,\,\, N \in \mathbb{N}  ~.\label{ppl}
\end{eqnarray}
where $\ell _{pl}$ is the Planck length.
When one moves away from the extremality and super-symmetry, the area products are discretized \cite{visser2} in
terms of the Planck area and the fine structure constant i.e.
\begin{eqnarray}
{\cal A}_{+} {\cal A}_{-} &=& \left(8\pi {\ell _{pl}}^2\right)^2 \left[\ell(\ell+1)+\frac{\alpha^2q^2}{4}\right]~.\label{pplfc}
\end{eqnarray}
which implies that the quantization rules are break down, only because the fine structure constant ($\alpha$) is not
an integer. Here $\alpha \in \mathbb{N}$ and $q\in \mathbb{N}$.

The fact is that, the Cauchy horizon is  an ``infinite blue-shift" region and classically
unstable due to the linear perturbation.  Thus, when an observer crosses the Cauchy
horizon $r=r_{-}$, he/she may speculated all of the events which occur at ``Region-I'' and
also seen the electromagnetic and gravitational field oscillations at infinite frequency
which is actually occur at finite frequency in the ``Region-I'' \cite{sch}.

Despite the above fact that the Cauchy horizon is an interesting venue where we  
study the following features of  both  the charged rotating space-time and the rotating space-time.
\begin{itemize}
\item We prove that in section (\ref{knewman}), like the area and entropy product,
the surface gravity product, surface temperature or black hole temperature product and
Komar energy product of both inner horizon and outer horizon do not shows any
global properties due to the mass dependence. Such products are \emph{not} universal 
in nature.

\item We explicitly show that in section (\ref{smar}), the black hole
mass or ADM mass can be expressed as in terms of the area of Cauchy horizon
${\cal H}^{-}$:
\begin{eqnarray}
{\cal M}^2 &=& \frac{{\cal A}_{-}}{16\pi}+\frac{4\pi J^2}{{\cal A}_{-}}
+\frac{Q^2}{2}+\frac{\pi Q^4}{{\cal A}_{-}} ~.\label{chy}
\end{eqnarray}
and we prove that the mass can be expressed as sum of  the surface energy, rotational energy and 
electro-magnetic energy of the Cauchy horizon ${\cal H}^{-}$:
\begin{eqnarray}
{\cal M} &=& {\cal E}_{s-}+ {\cal E}_{r-} + {\cal E}_{em-}
~. \label{summ}
\end{eqnarray}

\item Also we find that in section (\ref{ruffini}), the Christodoulou-Ruffini \cite{cr} mass formula may be 
expressed as in terms of the area of the  Cauchy horizon 
    ${\cal H}^{-}$:
\begin{eqnarray}
{\cal M}^2 &=& ({\cal M}_{irr -}+\frac{Q^2}{4 {\cal M}_{irr -} })^2+\frac{J^2}{4 ({\cal M}_{irr -})^2} ~. \label{irrCh}
\end{eqnarray}
\item We further investigate the laws of the black hole mechanics  for \emph{inner horizon} in section (\ref{laws}).
\item We also point out that the product of Christodoulou's irreducible mass of inner horizon (Cauchy horizon) and 
outer horizon (event horizon) are \emph{independent} of mass. i.e.
\begin{eqnarray}
{\cal M}_{irr+} {\cal M}_{irr-} &=& \sqrt{\frac{{\cal A}_{+} {\cal A}_{-}}{16\pi}}
                                 =  \sqrt{\frac{J^2+\frac{Q^4}{4}}{4}} ~. \label{irrmp}
\end{eqnarray}
\item We also shortly derive an identity $K_{\chi^{\mu}-}=2{\cal S}_{-} T_{-}$ on the
Cauchy horizon in section (\ref{komar1}).
\item The entropy of the Cauchy horizon may be expressed in the form ${\cal S}_{-}=\frac{E_{-}}{2T_{-}}$ is described
in section (\ref{komar}).
\end{itemize}

\section{\label{knewman} Charged Rotating Black Hole }

The complete gravitational  collapse of a charged body always produces Kerr-Newman (KN) black hole \cite{kn}, which is 
the most general class among the classical black hole solutions. It is also uniquely described by the electro 
vacuum black hole solutions of the Einstein-Maxwell system. It can be specified by three parameters: the black 
hole mass ${\cal M}$, the charge $Q$ and the angular momentum per unit mass $a=J/{\cal M}$. As long as 
${\cal M}^2 \geq Q^2 + a^2$ the KN metric describes a black hole, otherwise it has a naked ringlike singularity.
It possess two horizons namely event horizon (${\cal H}^+$) or outer horizon and Cauchy
horizon (${\cal H}^-$) or inner horizon. The proper radii of event horizon and Cauchy horizon are
\begin{eqnarray}
r_{\pm}&=& {\cal M}\pm\sqrt{{\cal M}^2-a^2-Q^2}\,\, \mbox{and}\,\,  r_{+}> r_{-}
\end{eqnarray}
whose product is
\begin{eqnarray}
r_{+} r_{-} &=& a^2+Q^2 ~.\label{prKN}
\end{eqnarray}
It is speculated that, it does not depend on mass but depends on charge and Kerr parameter \cite{kerr}.

Then the area\cite{Beken1,Beken2}  of both the horizons (${\cal H}^\pm$) are
\begin{eqnarray}
{\cal A}_{\pm} &=& \int\int \sqrt{g_{\theta\theta}g_{\phi\phi}} =4\pi(r_{\pm}^2+a^2) ~.\label{arKN}
\end{eqnarray}

The angular velocity of ${\cal H}^\pm$ are
\begin{eqnarray}
 {\Omega}_{\pm} &=& \frac{a}{r_{\pm}^2+a^2}  ~. \label{omega}
\end{eqnarray}

The semiclassical Bekenstein-Hawking entropy of ${\cal H}^\pm$ reads
(in units in which $G=\hbar=c=1$)

\begin{eqnarray}
{\cal S}_{\pm} &=& \frac{{\cal A}_{\pm}}{4} =\pi(r_{\pm}^2+a^2) ~.\label{etpKN}
\end{eqnarray}

The surface gravity of ${\cal H}^\pm$ is given by
\begin{eqnarray}
{\kappa}_{\pm} &=& =\frac{r_{\pm}-r_{\mp}}{2(r_{\pm}^2+a^2)} \,\, \mbox{and}\,\,  \kappa_{+}> \kappa_{-} ~.\label{sgKN}
\end{eqnarray}
and the black hole temperature or Hawking temperature of ${\cal H}^\pm$ reads as

\begin{eqnarray}
T_{\pm}&=&\frac{{\kappa}_{\pm}}{2\pi} =\frac{r_{\pm}-r_{\mp}}{4\pi (r_{\pm}^2+a^2)}  ~.\label{tmKN}
\end{eqnarray}
It should be noted that event horizon is hotter than the Cauchy horizon i.e. 
$T_{+} > T_{-} $.

The Komar energy for  ${\cal H}^\pm$ is given by (which will be discuss elaborately on the section (\ref{komar}))
\begin{eqnarray}
E_{\pm} &=& 2 {\cal S}_{\pm} T_{\pm}=\pm\sqrt{{\cal M}^2-a^2-Q^2} ~. \label{kengy}
\end{eqnarray}

Finally, the horizon Killing vector field may be defined for ${\cal H}^\pm$ are
\begin{eqnarray}
{\chi_{\pm}}^{a} &=& (\partial_{t})^a +\Omega_{\pm} (\partial_{\phi})^a~.\label{hkv}
\end{eqnarray}

If, in addition , the black hole is non-extremal 
(i.e., if there exists the trapped surface interior of the outer horizon) then the following relations are hold:
\begin{eqnarray}
{\cal A}_{+} &>& \sqrt{(8\pi)^2\left(J^2+\frac{Q^4}{4}\right)}> {\cal A}_{-}  ~.\label{prKN1}
\end{eqnarray}

Also the product of entropy is given by
\begin{eqnarray}
{\cal S}_{+} {\cal S}_{-} &=& (2\pi)^2\left(J^2+\frac{Q^4}{4}\right) ~.\label{pentropy}
\end{eqnarray}
It is also independent of mass $({\cal M})$.
The entropy of the non-extremal cases  satisfied the following inequality:
\begin{eqnarray}
{\cal S}_{+} & > & \sqrt{(2\pi)^2\left(J^2+\frac{Q^4}{4}\right)}> {\cal S}_{-}  ~.\label{KN entp}
\end{eqnarray}

Similarly,  we can compute the product of surface gravity of ${\cal H}^{\pm}$ is 
given by
\begin{eqnarray}
{\kappa}_{+} {\kappa}_{-} &=&- \frac{(r_{+}-r_{-})^2}{4(r_{+}^2+a^2)(r_{-}^2+a^2)} =
-\frac{{\cal M}^2-a^2-Q^2}{(r_{+}^2+a^2)(r_{-}^2+a^2)}~.\label{psg}
\end{eqnarray}

The product of surface temperature of ${\cal H}^{\pm}$ reads
\begin{eqnarray}
{T}_{+} {T}_{-} &=&- \frac{(r_{+}-r_{-})^2}{(4\pi)^2(r_{+}^2+a^2)(r_{-}^2+a^2)} =
-\frac{{\cal M}^2-a^2-Q^2}{(2\pi)^2(r_{+}^2+a^2)(r_{-}^2+a^2)}   ~.\label{pst}
\end{eqnarray}
and the product of Komar energy of ${\cal H}^{\pm}$ is
\begin{eqnarray}
E_{+}E_{-}  &=& (2{\cal S}_{+} T_{+}) (2 {\cal S}_{-} T_{-})=-({\cal M}^2-a^2-Q^2) ~. \label{kengy1}
\end{eqnarray}
It seems that these products  are \emph{not} universal.

In case of  pure Einstein gravity (with out Maxwell field),  the above relations are  reduces to:

 For the proper radii product of ${\cal H}^{\pm}$ :
\begin{eqnarray}
r_{+} r_{-} &=& a^2 ~.\label{prK}
\end{eqnarray}
 For the area product of ${\cal H}^{\pm}$ :
\begin{eqnarray}
{\cal A}_{+} {\cal A}_{-} &=& \left(8\pi J\right)^2 ~.\label{proarkerr}
\end{eqnarray}
 For the entropy product of ${\cal H}^{\pm}$ :
\begin{eqnarray}
{\cal S}_{+} {\cal S}_{-} &=& \left(2\pi J\right)^2~.\label{pentropy1}
\end{eqnarray}
 For the surface gravity product of ${\cal H}^{\pm}$ :
\begin{eqnarray}
{\kappa}_{+} {\kappa}_{-} &=&- \frac{(r_{+}-r_{-})^2}{4(r_{+}^2+a^2)(r_{-}^2+a^2)} =
-\frac{{\cal M}^2-a^2}{(r_{+}^2+a^2)(r_{-}^2+a^2)}~.\label{psgk}
\end{eqnarray}
 For the temperature product of ${\cal H}^{\pm}$  :
\begin{eqnarray}
{T}_{+} {T}_{-} &=&- \frac{(r_{+}-r_{-})^2}{(4\pi)^2(r_{+}^2+a^2)(r_{-}^2+a^2)} =
-\frac{{\cal M}^2-a^2}{(2\pi)^2(r_{+}^2+a^2)(r_{-}^2+a^2)}~.\label{pstk}
\end{eqnarray}

 For the Komar energy product of ${\cal H}^{\pm}$:
\begin{eqnarray}
E_{+}E_{-}  &=& (2{\cal S}_{+} T_{+}) (2 {\cal S}_{-} T_{-})=-({\cal M}^2-a^2) ~. \label{kengy2}
\end{eqnarray}

So, the product of the area and entropy of the both horizons are proportional to the square of the spin 
parameter $J$. Surface gravity product, surface temperature product and Komar energy product depends 
on mass. Thus,  we may conclude that they are \emph{not} universal
except the area product and entropy product.

\section{\label{smar} Smarr Formula for Cauchy horizon (${\cal H}^{-}$)}

In the original paper by Larry Smarr \cite{smarr},  the area for the charged rotating
black hole is described by the following relation
\begin{eqnarray}
{\cal A} &=& 4\pi \left( 2{\cal M}^2-Q^2 + 2\sqrt{{\cal M}^4-J^2-{\cal M}^2Q^2}\right) ~.\label{areaKN}
\end{eqnarray}
It is indeed constant over the exterior horizon. We suggest here that there are two horizons, so correspondingly 
both areas must be a constant i.e. the area can be
expressed as
\begin{eqnarray}
{\cal A}_{\pm} &=& 4\pi \left( 2{\cal M}^2-Q^2 \pm 2\sqrt{{\cal M}^4-J^2-{\cal M}^2Q^2}\right) ~.\label{areaKN2}
\end{eqnarray}
Inverting the above relation one can obtain the black hole mass or ADM mass can be
expressed as in terms of area of both horizons ${\cal H}^\pm$,
\begin{eqnarray}
{\cal M}^2 &=& \frac{{\cal A}_{\pm}}{16\pi}+\frac{4\pi J^2}{{\cal A}_{\pm}}
+\frac{Q^2}{2}+\frac{\pi Q^4}{{\cal A}_{\pm}} ~.\label{m2}
\end{eqnarray}
It is remarkable that the mass can be expressed as in terms of both area of
${\cal H}^+$ and ${\cal H}^-$.
Now what happens for the mass differential. It is indeed expressed as three
physical invariants of both ${\cal H}^+$ and ${\cal H}^-$,
\begin{eqnarray}
d{\cal M} &=& {\cal T}_{\pm} d{\cal A}_{\pm} + \Omega_{\pm} dJ +\Phi_{\pm}dQ
~. \label{dm}
\end{eqnarray}
where
\begin{eqnarray}
{\cal T}_{\pm} &=&  \frac{1}{{\cal M}} \left(\frac{1}{32 \pi}-\frac{2\pi J^2}{{\cal A}_{\pm}^2}-\frac{\pi Q^4}{2 {\cal A}_{\pm}^2} 
\right)\nonumber \\
\Omega_{\pm} &=& \frac{4\pi J}{{\cal M}{\cal A}_{\pm}} \nonumber\\
\Phi_{\pm} &=& \frac{1}{{\cal M}} \left(\frac{Q}{2}+\frac{2\pi Q^3}{{\cal A}_{\pm}} \right)  ~. \label{invar}
\end{eqnarray}
where
\begin{eqnarray}
{\cal T}_{\pm} &=& \mbox{Effective surface tension for ${\cal H}^+$ and ${\cal H}^-$} \nonumber \\
\Omega_{\pm} &=&  \mbox{Angular velocity for ${\cal H}^\pm$} \nonumber \\
\Phi_{\pm} &=& \mbox{Electromagnetic potentials for ${\cal H}^\pm$}\nonumber
\end{eqnarray}

The effective surface tension may be rewritten as
\begin{eqnarray}
{\cal T}_{\pm} &=& \frac{1}{{\cal M}} \left(\frac{1}{32 \pi}-\frac{2\pi J^2}{{\cal A}_{\pm}^2}-\frac{\pi Q^4}{2 {\cal A}_{\pm}^2} \right) \\
&=& \frac{1}{32 \pi {\cal M}} \left(1-\frac{16\pi^2(4J^2+Q^4)}{{\cal A}_{\pm}^2}\right)\nonumber\\
&=& \frac{1}{16 \pi {\cal M}} \left(1-\frac{(2{\cal M}^2-Q^2)}{r_{\pm}^2+a^2}\right)\nonumber\\
&=& \pm \frac{\sqrt{{\cal M}^2-a^2-Q^2}}{8\pi(r_{\pm}^2+a^2)} \nonumber\\
&=& \frac{r_{\pm}-{\cal M}}{8\pi(r_{\pm}^2+a^2)}= \frac{{\kappa}_{\pm}} {8\pi}
\end{eqnarray}
where $\kappa_{\pm}$ are the surface gravity of ${\cal H}^{\pm}$ as previously defined.

Thus the mass can be expressed in terms of these quantities both for ${\cal H}^\pm$ as a
simple bilinear form
\begin{eqnarray}
{\cal M} &=& 2{\cal T}_{\pm} {\cal A}_{\pm} + 2J \Omega_{\pm} +\Phi_{\pm} Q
~. \label{bilinear}
\end{eqnarray}
This has been derived from the homogenous function of degree $\frac{1}{2}$ in
$({\cal A}_{\pm}, J, Q^2)$.
Remarkably, ${\cal T}_{\pm}$,  $\Omega_{\pm}$ and $\Phi_{\pm}$ can be defined and
are constant on the ${\cal H}^+$ and ${\cal H}^-$ for any stationary, axisymmetric
space-time.
Since the $d{\cal M}$ is perfect differential, one may choose freely any path of
integration in $({\cal A}_{\pm}, J, Q)$ space. Thus the surface energy ${\cal E}_{s \pm}$
for ${\cal H}^+$ and ${\cal H}^{-}$ can be defined by
\begin{eqnarray}
{\cal E}_{s \pm} &=& \int_{0}^{{\cal A}_{\pm}} {\cal T} (\tilde{{\cal A}_{\pm}}
, 0 ,0) d\tilde{{\cal A}_{\pm}}; ~ \label{se}
\end{eqnarray}
the rotational energy  for ${\cal H}^+$ and ${\cal H}^{-}$ can be defined by
\begin{eqnarray}
{\cal E}_{r \pm} &=& \int_{0}^{J} \Omega_{\pm} ({\cal A}_{\pm}
, \tilde{J} ,0) d\tilde{J},\,\,  \mbox{${\cal A}_{\pm}$ fixed}; ~ \label{re}
\end{eqnarray}

and the electromagnetic energy  for ${\cal H}^+$ and ${\cal H}^{-}$ are
\begin{eqnarray}
{\cal E}_{em \pm} &=& \int_{0}^{Q} \Phi_{\pm} ({\cal A}_{\pm}
, J, \tilde{Q}) d\tilde{Q},\,\, \mbox{${\cal A}_{\pm}$, $J$  fixed}; ~ \label{re1}
\end{eqnarray}

We may rewritten the Eq. (\ref{bilinear}) as
\begin{eqnarray}
{\cal M} &=& \frac{{\kappa}_{\pm}}{4\pi}{\cal A}_{\pm} + 2J\Omega_{\pm}+\Phi_{\pm}Q
~. \label{bilinear1}
\end{eqnarray}
or
\begin{eqnarray}
{\cal M}-2J\Omega_{\pm}-\Phi_{\pm}Q &=& \frac{{\kappa}_{\pm}}{4\pi}{\cal A}_{\pm}
~. \label{bilinear2}
\end{eqnarray}
or
\begin{eqnarray}
{\cal M}-2J\Omega_{\pm}-\Phi_{\pm}Q &=& \frac{T_{\pm}}{2}{\cal A}_{\pm}
~. \label{bilinear3}
\end{eqnarray}
or
\begin{eqnarray}
\frac{{\cal M}}{2} &=& {T}_{\pm}{\cal S}_{\pm} + J\Omega_{\pm}+\frac{\Phi_{\pm}Q}{2}
~. \label{bilinear4}
\end{eqnarray}
This is recognize as a generalized \emph{Smarr-Gibbs-Duhem} relation for
${\cal H}^{\pm}$. Here `+' indicate for event horizon which was already
discussed in the literature\cite{davies}. We have derived the above relation
here for Cauchy horizon only and for our record we have also mentioned here
both the horizons.

Now defining a new parameter set $(\eta_{\pm}, \beta_{\pm}, \epsilon_{\pm})$
which is related to the quantities $({\cal A}_{\pm}, J, Q)$ to study the
intrinsic geometry for the Cauchy horizon ${\cal H}^{-}$ of a charged rotating black hole is given by
\begin{eqnarray}
\eta_{\pm} &=& \sqrt{r_{\pm}^2+a^2}=\sqrt{\frac{{\cal A}_{\pm}}{4\pi}} ~. \label{eta1}\\
\beta_{\pm} &=& \frac{a}{\sqrt{r_{\pm}^2+a^2}}=\frac{a}{\eta_{\pm}} ~. \label{beta1} \\
\epsilon_{\pm} &=& \frac{Q}{\eta_{\pm}} ~. \label{epsil}
\end{eqnarray}
Therefore the integrated mass formula for ${\cal H}^{\pm}$ is found to be

\begin{eqnarray}
{\cal M} &=& \frac{\eta_{\pm}(1+{\epsilon_{\pm}}^2)}{2\sqrt{1-{\beta_{\pm}}^2}}
~. \label{maseta}\\
{\cal E}_{s \pm}  &=& \frac{\eta_{\pm}}{2}  ~. \label{eta2}\\
{\cal E}_{r \pm}  &=& \frac{\eta_{\pm}}{2}[\frac{1}{\sqrt{1-{\beta_{\pm}}^2}}-1]
~. \label{beta2} \\
{\cal E}_{em \pm} &=& \frac{\eta_{\pm}{\epsilon_{\pm}}^2}{2\sqrt{1-{\beta_{\pm}}^2}}   ~. \label{epsil2}
\end{eqnarray}
with
\begin{eqnarray}
{\cal M} &=& {\cal E}_{s\pm}+ {\cal E}_{r\pm} + {\cal E}_{em\pm}
~. \label{summass}
\end{eqnarray}
Interestingly, mass can be expressed as sum of surface energy, rotational energy
and electro-magnetic energy of both the horizons ${\cal H}^{\pm}$.
We have already seen the  above discussion in \cite{smarr} for event horizon. We derived here
the above relation for Cauchy horizon only and for the sake of completeness
we have mentioned both the cases.

\section{\label{ruffini} Christodoulou's Irreducible Mass for Cauchy horizon}

Besides the black hole event horizon, there exist a second horizon inside the
black hole - the Cauchy horizon or inner horizon $({\cal H}^-)$. It is defined as the
future boundary of the domain of dependence of the $({\cal H}^+)$. What happens the
Christodoulou-Ruffini \cite{cr} mass formula for Cauchy horizon. This is an important
issue that we will discuss now.

The irreducible mass can be defined as
\begin{eqnarray}
 {\cal M}_{irr \pm} &=& \frac{\sqrt{r_{\pm}^2+a^2}}{2}=\sqrt{\frac{{\cal A}_{\pm}}{16\pi}}
~. \label{irrm}
\end{eqnarray}
where $+$ indicate for ${\cal H}^+$ and $-$ indicate for ${\cal H}^-$.
The area and angular velocity can be expressed as in terms of ${\cal M}_{irr \pm}$ :
\begin{eqnarray}
 {\cal A}_{\pm} &=& 16 \pi ({\cal M}_{irr \pm})^2=4 \pi \rho_{\pm}^2
~. \label{irrma}
\end{eqnarray}

and
\begin{eqnarray}
 {\Omega}_{\pm} &=& \frac{a}{r_{\pm}^2+a^2} = \frac{a}{4({\cal M}_{irr \pm})^2}
~. \label{iromega}
\end{eqnarray}
Interestingly, the product of the irreducible mass of  ${\cal H}^\pm$ are \emph {universal}.
\begin{eqnarray}
 {\cal M}_{irr+} {\cal M}_{irr-} &=& \sqrt{\frac{{\cal A}_{+} {\cal A}_{-}}{16\pi}} \\
                                 &=&  \sqrt{\frac{J^2+\frac{Q^4}{4}}{4}} ~. \label{irrmp1}
\end{eqnarray}

The rest mass of a rotating charged black hole is defined by the Christodoulou-Ruffini
mass formula (which may be expressed as in term of its irreducible mass,  angular momentum $J$  
and charge $Q$ ) reads as:
\begin{eqnarray}
{\cal M}^2 &=& ({\cal M}_{irr \pm}+\frac{Q^2}{4 {\cal M}_{irr \pm} })^2+\frac{J^2}{4 ({\cal M}_{irr \pm})^2} ~. \label{irrma1}\\
           &=& ({\cal M}_{irr \pm}+\frac{Q^2}{2 \rho _{ \pm} })^2+(\frac{J}{\rho}_{\pm})^2\\
           &=& (m_{r \pm})^2+(p_{\pm})^2\\
           &=& (\gamma_{\pm} m_{r \pm})^2 ~. \label{irrma2}
\end{eqnarray}
where $p_{\pm}=\frac{J}{{\rho}_{\pm}}$ is an effective momentum, and the effective rest mass $m_{r \pm}$
can be defined as

\begin{eqnarray}
 m_{r \pm} &=& {\cal M}_{irr \pm}+\frac{Q^2}{2\rho_{\pm}}={\cal M}_{irr \pm}+\frac{Q^2}{4 {\cal M}_{irr \pm}}
~. \label{kir4}
\end{eqnarray}
Also the corresponding gamma factor is given by
\begin{eqnarray}
 \gamma_{\pm} &=& \frac{1}{\sqrt{1-v_{\pm}^2}}=\frac{1}{\sqrt{1-\frac{a^2}{4 ({\cal M}_{irr \pm})^2}}}~. \label{kir6}
\end{eqnarray}

It may be noted that $(v_{\pm})^2$ is a strange product of the angular velocity and angular momentum
(the conjugate momentum variable) divided by the mass of the black hole.

\begin{eqnarray}
 (v_{\pm})^2 &=& a \Omega_{\pm}=\frac{J}{M} \Omega_{\pm} ~. \label{kir7}
\end{eqnarray}
One may compare various formulas for the black hole space-time and certain formulas from
mechanics and electromagnetism by rewritten the formula either in ${\cal M}_{irr \pm}$ or $\rho_{\pm}$
in geometric units which corresponds to a mass or length variable.

A reversible process is characterized by an unchanged irreducible mass. Whereas
an irreversible process is characterized by an increase in irreducible mass of a
black hole. It should be noted that there exist no process which will decrease the
${\cal M}_{irr}$ for Cauchy horizon.

This mass decompose into irreducible mass ${\cal M}_{irr}$ and a rotational energy
${\cal M}-{\cal M}_{irr}$ for Kerr black hole as shown by Christodoulou \cite{cd}.
For Kerr-Newman black hole the mass can be written as both for ${\cal H}^+$ and ${\cal H}^-$ in
the small angular momentum limit, one has
\begin{eqnarray}
{\cal M} &=& {\cal M}_{irr \pm}+\frac{Q^2}{4 {\cal M}_{irr \pm}} +{\cal K}_{\pm}
~. \label{kir}
\end{eqnarray}

where,
\begin{eqnarray}
{\cal K}_{\pm} &=& \frac{1}{2} \left({\cal M}_{irr \pm}+\frac{Q^2}{4 {\cal M}_{irr \pm}}\right) \left( v_{\pm} \right)^2
~. \label{kir1}
\end{eqnarray}
looks like an expression for kinetic energy in classical mechanics.

The effective speed $v_{\pm}$ is given by
\begin{eqnarray}
 v_{\pm} &=& \frac{{p}_{\pm}}{{\cal M}} =\frac{J}{{\cal M} \rho_{\pm} }=
 \frac{a}{\rho_{\pm}}=\frac{a}{2 {\cal M}_{irr \pm}}=\rho_{\pm}{\Omega}_{\pm}
~. \label{kir2}
\end{eqnarray}
where, $\rho_{\pm}=\sqrt{r_{\pm}^2+a^2}=2 {\cal M}_{irr \pm}$. Thus the equation (\ref{kir1}) reduces to

\begin{eqnarray}
{\cal K}_{\pm} &=& \frac{1}{2} m_{r \pm} \left( v_{ \pm} \right)^2 =\frac{J^2}{2I_{\pm}}
=\frac{1}{2} I_{\pm} {{\Omega}_{\pm}}^2
~. \label{kir3}
\end{eqnarray}
where, $I_{\pm}=m_{r \pm} \left( \rho_{ \pm} \right)^2 $  playing the role of
a moment of inertia in this limit. The above discussion  for event horizon can be found 
in \cite{bjr}. We have derived the above formula here for CH only.  

When the Penrose process \cite{penrose1,penrose2} are taking into account, which led to the following exact
differential relationship between mass and angular momentum which also characterized by reversible transformations, as
described by Christodoulou  and Ruffini

\begin{eqnarray}
d{\cal M} &=& \frac{a dJ+r_{\pm} Q dQ}{r_{\pm}^2+a^2}~. \label{crmass}
\end{eqnarray}
After integration we obtain the Christodoulou  and Ruffini mass formula (\ref{irrma1}),
when the following condition is satisfied:
\begin{eqnarray}
\frac{J^2}{4({\cal M}_{irr \pm})^2} +\frac{Q^4}{16({\cal M}_{irr \pm})^4} \leq 1 \label{crm1}
\end{eqnarray}

\section{\label{laws} The Four Laws of Black Hole Mechanics on the EH $({\cal H}^{+})$ and CH $({\cal H}^{-})$}

Following the remarkable discovery by Carter, Hawking and Bardeen \cite{bcw}, we reformulate the black hole thermodynamics
both for the event horizon and the Cauchy horizon which is analogous to the classical laws of thermodynamics as follows:
\begin{itemize}
\item The Zeroth Law: The surface gravity, $\kappa_{\pm}$ of a stationary black hole is constant over both the
event horizon (${\cal H }^{+}$) and Cauchy horizon (${\cal H }^{-}$) respectively.

\item The First Law: Any perturbation of a stationary black holes, the change of mass (change of energy) is
related to change of mass, angular momentum, and electric charge by:

\begin{eqnarray}
d{\cal M} &=& \frac{{\kappa}_{\pm}} {8\pi} d{\cal A}_{\pm} + \Omega_{\pm} dJ +\Phi_{\pm}dQ ~. \label{dm1}
\end{eqnarray}
It can be seen that $\frac{{\kappa}_{\pm}} {8\pi}$ is analogous to the temperature of ${\cal H}^{\pm}$  in the same way that
${\cal A}_{\pm}$ is analogous to entropy. It should be noted that $\frac{{\kappa}_{\pm}} {8\pi}$ and ${\cal A}_{\pm}$
are distinct from the temperature and entropy of the black hole.

The above expression $\frac{{\kappa}_{\pm}} {8\pi}$ can be derived from the Eq.
(\ref {invar}) in the following way.

The effective surface tension can be rewritten as
\begin{eqnarray}
{\cal T}_{\pm} &=&  \frac{{\kappa}_{\pm}} {8\pi}
= \frac{\partial {\cal M}}{\partial {\cal A}_{\pm}}
\end{eqnarray}
and
\begin{eqnarray}
\Omega_{\pm} &=& \frac{4\pi J}{{\cal M}{\cal A}_{\pm}}=\frac{\partial {\cal M}}{\partial J} \\
\Phi_{\pm} &=& \frac{1}{{\cal M}} \left(\frac{Q}{2}+\frac{2\pi Q^3}{{\cal A}_{\pm}}  \right) =
\frac{\partial {\cal M}}{\partial Q} ~. \label{invar1}
\end{eqnarray}
\item The Second Law: The area  ${\cal A}_{\pm}$ of both event horizons $({\cal H}^{+})$ and
 Cauchy horizons $({\cal H}^{-})$ never decreases, i.e.
\begin{eqnarray}
 d{\cal A}_{\pm} &=& \frac{4 {\cal A}_{\pm}}{r_{\pm}-r_{\mp}} \left(d{\cal M}-\vec{\Omega}_{\pm}.d\vec{J}
-\Phi_{\pm} dQ \right)\geq 0 \label{arth}
\end{eqnarray}
or
\begin{eqnarray}
 d{\cal M}_{irr\pm}&=& \frac{2 {\cal M}_{irr\pm}}{r_{\pm}-r_{\mp}} \left(d{\cal M}-\vec{\Omega}_{\pm}.d\vec{J}
-\Phi_{\pm} dQ \right) \geq 0 \label{arth1}
\end{eqnarray}

The change in irreducible mass of both event horizons $({\cal H}^{+})$ and Cauchy horizons $({\cal H}^{-})$ can never be negative.

It follows that immediately from the above equation
\begin{eqnarray}
  d{\cal M} > \vec{\Omega}_{\pm}.d\vec{J}+\Phi_{\pm} dQ  \label{srad}
\end{eqnarray}

\item The Third Law: It is impossible by any mechanism, no matter how idealized, to reduce, $\kappa_{\pm}$ the surface
gravity of  both event horizon $({\cal H}^{+})$ and Cauchy horizon $({\cal H}^{-})$ to zero by a finite sequence of
operations.
\end{itemize}

\section{\label{komar1} $\mbox{Komar Conserved Quantity}=2\times\mbox{Entropy on }{\cal H}^{\pm}\times \mbox {Temperature on}{\cal H}^{\pm} $ \,\, or $K_{\chi^{\mu}\pm}=2{\cal S}_{\pm} T_{\pm}$ }

It is known \cite{modak} that on the ${\cal H}^{+}$ the Komar conserved quantity $(K_{\chi^{\mu}+})$ corresponding 
to a null Killing vector ${\chi^{\mu}}_+$ is equal to the twice of the product of entropy (${\cal S}_{+}$) on
${\cal H}^{+}$ and temperature ($T_{+}$) on ${\cal H}^{+}$. Here, we shall derive the similar expression which exists 
on ${\cal H}^{-}$.
Thus, we have to prove the identity $K_{\chi^{\mu}-}=2 {\cal S}_{-} T_{-}$ on the ${\cal H}^{-}$.

Due to the stationarity and axially symmetric nature of Kerr Newman space-time, the space-time has two Killing vectors.
These two vectors are ${\xi^\mu}_{(t)}=(1,0,0,0)$ and  ${\xi^\mu}_{(\phi)}=(0,0,0,1)$ which corresponds to timelike and spacelike
at the asymptotic limit. Thus we can define a Killing vector on both the $({\cal H}^{\pm})$ which is the combination
of these two vectors.

\begin{eqnarray}
{\chi^\mu}_{\pm} &=& {\xi^\mu}_{(t)} + \Omega_{\pm} {\xi^\mu}_{(\phi)} =(1,0,0,\Omega_{\pm}) ~. \label{kv}
\end{eqnarray}
It should be noted that on the $({\cal H}^{\pm})$, $\chi ^{\mu} \chi_{\mu}|_{r=r_{\pm}} =0$, then $\xi^{\mu}$ becomes
a null Killing vector.
Now, we can define a Komar conserved quantity on $({\cal H}^{\pm})$ which corresponds to the Killing
vectors are given by
\begin{eqnarray}
K_{\chi^\mu \pm} &=&K_{{\xi^\mu}_{(t)}} + \Omega_{\pm} K_{{\xi^\mu}_{(\phi)}} ~.\label{kv1}
\end{eqnarray}
where,  $K_{\chi^\mu \pm}$ is the Komar conserved quantity corresponding to the timelike Killing vector
which may be  defined as
\begin{eqnarray}
K_{{\xi^\mu}_{(t)}} &=& -\frac{1}{8\pi}  \int _{\partial \Sigma} \ast d\sigma ~. \label{kv2}
\end{eqnarray}
whose one form is given by
\begin{eqnarray}
\sigma = \xi_{(t)\mu}dx^{\mu}=g_{t\mu}dx^{\mu}=g_{tt} dt +g_{t\phi} d\phi ~. \label{kv3}
\end{eqnarray}
and $\ast d\sigma$ is the dual to the two form $d\sigma$. $d\Sigma$ is
an appropriate boundary surface of a spatial three volume ($\Sigma$).

Similarly, we can define Komar conserved quantity as $({\cal H}^\pm)$  correspond to the
spacelike Killing vector is given by
\begin{eqnarray}
K_{{\xi^\mu}_{(\phi)}} &=& -\frac{1}{8\pi}  \int _{\partial \Sigma} \ast d\eta ~. \label{kv4}
\end{eqnarray}
where the spacelike killing one form is defined as
\begin{eqnarray}
\eta &=& \xi_{(\phi)\mu}dx^{\mu}=g_{\mu\phi}dx^{\mu}=g_{t\phi} dt +g_{\phi\phi} d\phi ~. \label{kv5}
\end{eqnarray}
After some computations,  we find
\begin{eqnarray}
K_{\chi^{\mu}\pm} &=& {\cal M} -2{\cal M}\frac{a^2}{r_{\pm}^2+a^2}- \frac{Q^2 r_{\pm}}{r_{\pm}^2+a^2} ~. \label{kv6}
\end{eqnarray}
Using the expressions $\Omega_{\pm}$ and $r_{\pm}$, we simplify the above equation to rewritten in the
compact form
\begin{eqnarray}
K_{\chi^{\mu}\pm} &=& \pm \sqrt{{\cal M}^2-a^2-Q^2}= \frac{r_{\pm}-r_{\mp}}{2} \\
                  &=& 2 \left[\pi (r_{\pm}^2+a^2)\right] \frac{r_{\pm}-r_{\mp}}{4\pi (r_{\pm}^2+a^2)} \\
                  &=& \frac{{\cal A}_{\pm}\kappa_{\pm}}{4\pi} =2 \left(\frac{{\cal A}_{\pm}}{4}\right) \left(\frac{\kappa_{}\pm }{2\pi}\right) \\
                  &=& 2 {\cal S}_{\pm} T_{\pm}   ~. \label{kv7}
\end{eqnarray}
Thus we have obtained on the $({\cal H}^\pm)$ the Komar conserved charge corresponding to the null Killing vector is twice the product of the entropy and the surface temperature of the Kerr Newman black hole.
We can connect this quantity with a similar relation which has been derived in the previous section on the
$({\cal H}^\pm)$.
\begin{eqnarray}
E_{\pm} &=&  2 {\cal S}_{\pm} T_{\pm}   ~. \label{kv8}
\end{eqnarray}
where $E_{\pm}$ is the Noether charge of diffeomorphism symmetry on $({\cal H}^\pm)$. It should be noted that the above 
relation has been derived particularly for a static local Killing horizon \cite{paddy} which may or may not be a black 
hole event horizon. However Kerr Newman space-time is a stationary and therefore does it valid for any stationary space-time
which is not clear, but we have explicitly compute a similar relation on the Cauchy horizon ${\cal H}^{-}$. Introducing 
the scalar potential on the $({\cal H}^\pm)$,  we get
\begin{eqnarray}
 \Phi_{\pm} &=& \frac{Qr_{\pm}}{r_{\pm}^2+a^2} ~. \label{kv9}
\end{eqnarray}

Eq. (\ref{kv4}) can be rewritten in the form

\begin{eqnarray}
K_{\chi^{\mu}\pm} &=& {\cal M} -2J\Omega_{\pm}-\Phi_{\pm}Q  ~. \label{kv10}
\end{eqnarray}
Again from the Eq.(\ref{bilinear}) we have

\begin{eqnarray}
{\cal M} -2J\Omega_{\pm}-\Phi_{\pm}Q &=&  \frac{{\cal A}_{\pm}\kappa_{\pm}}{4\pi} \\
                                     &=&   \frac{{\cal A}_{\pm}T_{\pm}}{2}~. \label{kv11}
\end{eqnarray}

This is the well known Smarr formula on the $({\cal H}^\pm)$.  In the literature,  we have seen the above 
discussion for the event horizon only. We extend the above relations, particularly for the Cauchy horizon.
For the completeness, we  compute all the relations or formula for the event horizon also.

\section{\label{komar} Generalized Smarr Formula for Mass on the Cauchy horizon}

In this section,  we will derive for a stationary state black hole  space-time  the entropy can be expressed
as ${\cal S}_{\pm}=\frac{E_{\pm}}{2T_{\pm}} $ on the $({\cal H}^{\pm})$, where $T_{\pm}$ is the Hawking
temperature on $({\cal H}^{\pm})$ and $E_{\pm}$ is shown to be the Komar energy on the $({\cal H}^{\pm})$.
We also derive the generalized Smarr formula for mass on the $({\cal H}^{\pm})$. Which is compatible with the 
relation in Eq. (\ref{kv11}).
We have already known from \cite{rabin},  the Komar energy for Kerr Newman black hole in a compact 
form on the $({\cal H}^{+})$  is given by 

\begin{eqnarray}
2 {\cal S}_{+} T_{+} &=& E_{+}=\sqrt{{\cal M}^2-a^2-Q^2}\\
&=& {\cal M}-\frac{Q^2}{r_{+}}-2J\Omega_{+}\left(1-\frac{Q^2}{2{\cal M}r_{+}} \right)\\
&=& {\cal M}-2J\Omega_{+}-QV_{+} ~. \label{st1}
\end{eqnarray}
where $V_{+}=\frac{Q}{r_{+}}-\frac{JQ\Omega_{+}}{{\cal M}r_{+}}$.
Similarly, we can obtain easily the Komar energy on the Cauchy horizon for Kerr Newman black hole:

\begin{eqnarray}
2 {\cal S}_{-} T_{-}=E_{-}=-\sqrt{{\cal M}^2-a^2-Q^2} \label{st2}
\end{eqnarray}
Remarkably,  the energy is negative which also reverify that the Killing vector field is negative inside
the ${\cal H}^+$. Thus the energy is negative on the ${\cal H}^-$ due to this fact.

Again,  we can rewrite the Eq. (\ref{st2}) as

\begin{eqnarray}
2 {\cal S}_{-} T_{-}=E_{-}=-\sqrt{{\cal M}^2-a^2-Q^2}= {\cal M}-2J\Omega_{-}-QV_{-} ~. \label{st3}
\end{eqnarray}
where,  $V_{-}=\frac{Q}{r_{-}}-\frac{JQ\Omega_{-}}{{\cal M}r_{-}}$.

Thus Eq. (\ref{st1}) and  Eq. (\ref{st3}) can be rewritten for both  the horizons $({\cal H}^\pm)$ in a
compact form as
\begin{eqnarray}
2 {\cal S}_{\pm} T_{\pm}=E_{\pm}=\pm\sqrt{{\cal M}^2-a^2-Q^2}= {\cal M}-2J\Omega_{\pm}-QV_{\pm} ~. \label{st4}
\end{eqnarray}

\section{\label{extremal} Degenerate Black Hole or Extremal Black Hole}
Thus one may define an extremal black hole is a black hole, when the radius of 
event horizons and Cauchy horizons are converging i.e.,
\begin{eqnarray}
r_{+} &=& r_{-} \label{a0}
\end{eqnarray}
or, when the area of two horizons are merging i.e.,
\begin{eqnarray}
{\cal A}_{+} &=& {\cal A}_{-} \label{a1}
\end{eqnarray}
or,
when the entropy of two horizons are coincident i.e.,
\begin{eqnarray}
{\cal S}_{+} &=& {\cal S}_{-} \label{s1}
\end{eqnarray}
or,
when the surface gravity of both horizons are equal i.e.,
\begin{eqnarray}
\kappa_{+} &=& \kappa_{-} \label{k1}
\end{eqnarray}
or,
when the temperature of both horizons are same i.e.,
\begin{eqnarray}
{ T}_{+} &=& { T}_{-} \label{t1}
\end{eqnarray}
or,
when the angular velocity of both horizons are coincident i.e.,
\begin{eqnarray}
{\Omega}_{+} &=& {\Omega}_{-} \label{m1}
\end{eqnarray}
or,
when the irreducible mass of both horizons are equal i.e.,
\begin{eqnarray}
{\cal M}_{irr+} &=& {\cal M}_{irr-} \label{irm}
\end{eqnarray}
If any one of the above properties are satisfied then a black hole is said to be
an extremal black hole.
Thus one gets the area in the extremal limit
\begin{eqnarray}
{\cal A}_{+} &=& {\cal A}_{-} = 8 \pi \sqrt {J^2+\frac{Q^4}{4}} \label{a11}
\end{eqnarray}
As a result of (\ref{a11}), the another relation
\begin{eqnarray}
\frac{J^2}{{\cal M}^2} +Q^2 &=& {{\cal M}_{CR}}^2 \label{a10}
\end{eqnarray}
of the extremal KN space-time continues to be hold in presence of the surrounding matter
in accordance with the fact that KN black holes are degenerate and  if they are extremal.
${\cal M}_{CR}$ denotes Christodoulou and Ruffini mass. Also for the entropy one obtains,
\begin{eqnarray}
{\cal S}_{+} &=& {\cal S}_{-} = 2 \pi \sqrt {J^2+\frac{Q^4}{4}} \label{a12}
\end{eqnarray}
It is well known that the surface gravity and surface temperature  goes
to zero at the extremal limit i.e. $\kappa_{+} = \kappa_{-}=0$ and
${ T}_{+} ={ T}_{-}=0$.

It may be noted that the Komar energy goes to zero at the extremal limit
i.e.  $E_{+}=E_{-}=0$. This may implies that this is the another way to seeing the
\emph{discontinuity} between extremal space-time and non-extremal space-time.

\section{Discussions}

In this work, we have derived the Smarr formula on the Cauchy horizon $({\cal H}^{-})$.
We have proposed the four laws of black hole mechanics for inner horizon $({\cal H}^{-})$. We have found, in contrast 
to some earlier work \cite{castro,det} particularly for the first law of inner horizon $({\cal H}^{-})$, a complete 
consistency between our results and their results.

We have also demonstrated that the area product, horizon radii product, entropy product and irreducible mass product 
of the event horizon and Cauchy horizon although are universal, the surface gravity product, surface temperature product
and Komar energy product are not universal for Kerr-Newman black hole.

We have also defined the Christodoulou and Ruffini mass on the Cauchy horizon.  We have further showed that the 
identity $K_{\chi^{\mu}-}=2 {\cal S}_{-} T_{-}$ is valid on the inner horizon (${\cal H}^{-}$) and also the 
Komar energy in a compact form $E_{-}=2 {\cal S}_{-} T_{-}$. Which also relates the generalized  
Smarr formula $ E_{-}= {\cal M}-2J\Omega_{-}-QV_{-}$ on  the  ${\cal H}^{-}$.

Another interesting point we have found that the Komar energy goes to zero at the
extremal limit indicates a discontinuity between extremal space-time and 
non-extremal space-time.


\begin{thebibliography}{99}

\bibitem{smarr} L. Smarr, \textit{ Phys. Rev. Lett.}  {\bf 30} 71 (1973); {\bf 31}, 521 (1973).

\bibitem{ansorg} M. Ansorg and J. Hennig,  \textit{ Classical Quantum Gravity} {\bf 25} {222001} (2008).

\bibitem{ansorg1} M. Ansorg and J. Hennig, \textit{ Phys. Rev. Lett.}  {\bf 102} {221102} (2009).

\bibitem{ansorg2} M. Ansorg, J. Hennig and C. Cederbaum, \textit {Gen. Rel. Grav.}  {\bf 43} {1205} (2011).

\bibitem{larsen} F. Larsen, \textit{Phys. Rev.}  {\bf D 56} {1005} (1997).

\bibitem{cvetic} M. Cvetic, G. W. Gibbons and C. N. Pope, \textit{ Phys. Rev. Lett.}  {\bf 106} {121301} (2011).

\bibitem{cvetic1} M. Cvetic and F. Larsen, \textit {J. High Energy Phys.} {\bf 09} {088} (2009).

\bibitem{chen} B.~Chen, S.~X.~Liu and J.~J.~Zhang, \textit{ J. High Energy Phys.}  {\bf 017}, 1211 (2012).

\bibitem{castro} A. Castro and M. J. Rodriguez, \textit{Phys. Rev.} {\bf D 86} {024008} (2012).

\bibitem{visser1} M. Visser, \textit{ Phys. Rev.}  {\bf D 88} {044014} (2013).

\bibitem{val} V. Faraoni, A. F. Z. Moreno \textit {`` Are quantization rules for
horizon areas universal? ''}, arXiv: 1208.3814 [hep-th] (2013).

\bibitem{visser2} M. Visser, \textit{ J. High Energy Phys.}  {\bf 06} {023} (2012).


\bibitem{sch} S. Chandrashekar, {\it The Mathematical Theory of Black Holes}, Clarendon Press, Oxford (1983).


\bibitem{cr} D. Christodoulou and R. Ruffini, \textit{ Phys. Rev. D.}  {\bf 4} {3552} (1971).

\bibitem{kn} E. Newman, K. Chinnaparad, A. Exton, A. Prakash, R. Torrence,
\textit {J. Math. Phys.} {\bf 6}, 918-919 (1965).

\bibitem{kerr} R. P. Kerr, \textit{ Phys. Rev. Lett.}  {\bf 11}, 237, (1963).


\bibitem{Beken1} J. D. Bekenstein, \textit{Phys. Rev.} {\bf D 7} 2333 (1973).

\bibitem{Beken2} J. D. Bekenstein, \textit{ Phys. Rev.} {\bf D 9} 3292 (1974).

\bibitem{davies} P. C. W. Davies,  \textit{ Rep. Prog. Phys., Vol. 41 }, (1978).

\bibitem{cd} D. Christodoulou, \textit {Phys. Rev. Lett.} {\bf 25} {1596} (1970).

\bibitem{bjr} D. Bini, R. Jantzen, R. Ruffini, \textit{``Reinterpretation of the Mass Formula for Black Hole''} (2012).

\bibitem{penrose1} R. Penrose and R. M. Floyd, \textit{Nature} {\bf 229}, 177 (1971).

\bibitem{penrose2} R. Penrose, \textit{ Riv. Nuovo Cimento}  {\bf 1}, 252 (1969).

\bibitem{bcw} J. M. Bardeen, B. Carter, S. W. Hawking,  \textit{Commun. Math. Phys.} {\bf 31}, 161 (1973).

\bibitem{modak} S. Modak and S. Samanta, \textit{ Int. J. Theor. Phys.}  {\bf 51} (2012) 1416-1424.

\bibitem {paddy} T. Padmanabhan, \textit{ Class. Quant. Grav.} {\bf 21} (2004) 4485-4494.

\bibitem{rabin} R. Banerjee and B. R. Majhi, \textit {Phys. Rev. D.} {\bf 81} {124006,} (2010).
  
\bibitem{det} S. Detournay, \textit{``Inner Mechanics of 3d Black Holes''}, arXiv: 1204.6088 [hep-th] (2012).

\end{thebibliography}
\end{document}